# Phonon Anomalies, Orbital-Ordering and Electronic Raman Scattering in iron-pnictide $Ca(Fe_{0.97}Co_{0.03})_2As_2$: Temperature-dependent Raman Study

Pradeep Kumar[1], D. V. S. Muthu[1], L. Harnagea[2], S. Wurmehl[2], B. Büchner[2] and A. K. Sood[1,*]

[1]Department of Physics, Indian Institute of Science, Bangalore -560012, India

[2]Leibniz-Institute for Solid State and Materials Research, (IFW)-Dresden, D-01171 Dresden, Germany

## ABSTRACT

We report inelastic light scattering studies on $Ca(Fe_{0.97}Co_{0.03})_2As_2$ in a wide spectral range of 120-5200 cm$^{-1}$ from 5K to 300K, covering the tetragonal to orthorhombic structural transition as well as magnetic transition at $T_{sm}$ ~ 160K. The mode frequencies of two first-order Raman modes $B_{1g}$ and $E_g$, both involving displacement of Fe atoms, show sharp increase below $T_{sm}$. Concomitantly, the linewidths of all the first-order Raman modes show anomalous broadening below $T_{sm}$, attributed to strong spin-phonon coupling. The high frequency modes observed between 400-1200 cm$^{-1}$ are attributed to the electronic Raman scattering involving the crystal field levels of $d$-orbitals of $Fe^{2+}$. The splitting between $xz$ and $yz$ $d$-orbital levels is shown to be ~ 25 meV which increases as temperature decreases below $T_{sm}$. A broad Raman band observed at ~ 3200 cm$^{-1}$ is assigned to two-magnon excitation of the itinerant Fe 3d antiferromagnet.

## 1. INTRODUCTION

The discovery of superconductivity in LaFeAsO$_{1-x}$F$_x$ [1] has generated enormous interest to investigate these systems experimentally as well as theoretically. Till now, various class of iron-based superconductors (FeBS) have been discovered such as AFe$_2$As$_2$ (A = Ba, Ca, Sr) [2], LiFeAs [3], Fe$_{1+y}$Te$_{1-x}$Se$_x$ [4], Ba$_4$Sc$_3$O$_{7.5}$Fe$_2$As$_2$ and Ca$_4$Al$_2$O$_{6-x}$Fe$_2$As$_2$ [5-6]. However, the origin of the pairing mechanism for superconductivity is still elusive. The parent compound exhibits long-range anti-ferromagnetic (AFM) ordering, which is suppressed on doping and subsequently superconductivity emerges [2,7]. Earlier theoretical studies suggested that AFM spin-fluctuations could induce electron pairing [8]. However, recent experimental as well theoretical studies have argued that coupled orbital and spin ordering can play an important role in the superconducting pairing mechanism [9-15]. There is a growing body of evidence that intricate interplay between spin (magnon) and orbitals (orbiton) degrees of freedom leads to novel electronic phases in FeBS.

Strongly electron correlated systems such as manganites, ruthenates and high temperature superconductors have been shown to support orbital excitations (orbitons), which are invoked to explain their interesting behavior [16-20]. The partially filled $d$-electron subshell ($d_{yz/xz}$) of Fe$^{2+}$ in FeBS can have orbital excitations [10-15] due to breaking of rotational symmetry of the $d_{xz}$ and $d_{yz}$ orbitals. Since the Fermi surface (FS) in FeBS is mainly composed of different components of the t$_{2g}$ orbitals of Fe$^{2+}$, namely $d_{xy}$, $d_{xz}$ and $d_{yz}$, these orbitals are very important in the paring mechanism. The $d_{xz}$ and $d_{yz}$ bands have strong in-plane anisotropy and play an important role in the electronic ordering observed in Ca(Fe$_{1-x}$Co$_x$)$_2$As$_2$ [20], similar to ruthenates [21] and this nematic ordering has been interpreted as orbital ordering between $d_{xz}$ and $d_{yz}$ orbitals [18, 20]. Therefore, the orbital excitations may be crucial for the coupling between the electrons which allow them to flow without resistance, suggesting an unconventional origin of superconductivity mediated by orbital and spin fluctuations [8-15].

Towards understanding the elementary excitations in FeBS, a few temperature-dependent Raman studies have been reported on these iron-based superconductors "1111" [22-23], "122" [24-31], "11" [32-34], "42622" [35] systems. The phonon anomalies observed in these systems have been attributed to mechanism such as strong spin-phonon coupling, opening of the spin density wave (SDW) gap, first-order structural phase transition and change of the charge distribution within the Fe-As plane. There have been reports of Raman studies on $Ca(Fe_{1-x}Co_x)_2As_2$ dealing with first-order phonons [25, 31], where the anomalies in the phonon frequencies and linewidths have been attributed to strong spin-phonon coupling and opening of the spin density wave gap below $T_{sm}$. We note that there are a few reports where the broad band observed in $Fe_{1+y}Se_xTe_{1-x}$ and $BaFe_{2-x}Co_xAs_2$ centered near ~ 2200 cm$^{-1}$ has been attributed to the two-magnon Raman scattering [29-30, 33]. In this paper, we study $Ca(Fe_{0.97}Co_{0.03})_2As_2$ having $T_{sm}$ ~ 160K [36], focusing on the phonon anomalies, the broad high energy excitation observed near 3200 cm$^{-1}$ and electronic excitations associated with the crystal field split $d$-level of $Fe^{2+}$. The broad mode at ~ 3200 cm$^{-1}$ is assigned to the two-magnon excitation. In addition, all the first-order modes show anomalous temperature dependence below $T_{sm}$ attributed to strong spin-phonon coupling.

## 2. EXPERIMENTAL DETAILS

Single crystals of $Ca(Fe_{0.97}Co_{0.03})_2As_2$ were prepared and characterized as described in ref. 36. Unpolarised micro-Raman measurements were performed in backscattering geometry using $\lambda_L$ = 514.5 nm line of an Ar-ion Laser (Coherent Innova 300) and Raman spectrometer (DILOR XY) coupled to a liquid nitrogen cooled CCD detector. The crystal surface facing the incident radiation is *ab* plane. Temperature variation was done from 5K to 300K with a temperature accuracy of ± 0.1K using continuous flow He cryostat (Oxford Instrument).

## 3. RESULTS AND DISCUSSIONS

### 3.1. Raman Scattering from Phonons

$CaFe_2As_2$ has a layered structure belonging to the tetragonal *I4/mmm* space group. There are four Raman active modes belonging to the irreducible representation $A_{1g}$ (As) + $B_{1g}$ (Fe) +$2E_g$ (As and Fe). Figure 1 shows Raman spectrum at 5K, revealing 7 modes labeled as S1 to S7 in the spectral range of 140-1200 $cm^{-1}$. Spectra are fitted to a sum of Lorentzian functions. The individual modes are shown by thin lines and the resultant fit by thick lines. Inset of figure shows the integrated intensity raio of mode S4 with respect to mode S3. Following the previous Raman studies on "122" systems, we assign the modes as S1 : 181 $cm^{-1}$ ($A_{1g}$ : As) ; S2 : 189 $cm^{-1}$ ($B_{1g}$ : Fe) and S3: 232 $cm^{-1}$ ($E_g$ : Fe and As). Figure 2 shows the spectra at a few typical temperatures. Figure 3 shows the temperature dependence of the mode frequencies (Panel-a) and their full width at half maxima (FWHM) (Panel-b). The solid lines are drawn as guide to the eyes. The following important observations can be made: (i) the frequencies of the modes S2 and S3 show abrupt hardening below $T_{sm}$, whereas mode S1 shows normal temperature dependence from 5K to 300K as expected due to anharmonic interactions. The dotted line for the mode S1 is a fit to a simple cubic anharmonicity model where the phonon decays into two phonons of equal frequency, giving temperature dependence of $\omega(T)$ as [37] $\omega(T) = \omega(0)+C[1+2n(\omega(0)/2)]$, where $n(\omega) = 1/(\exp(\hbar\omega/\kappa_B T) -1)$ is the Bose-Einstein mean occupation number and C is the self-energy parameter. Fitting parameters for the mode S1 are $\omega_0 = 181.5 \pm 0.2$ , $C = -0.66 \pm 0.1$ (ii) The FWHM decreases as temperature is decreased from 300K to $T_{sm}$ due to reduced anharmonicity. Interestingly, below $T_{sm}$, the linewidths of all the three phonon modes show anomalous broadening.

The anomalous hardening of modes S2 and S3 below $T_{sm}$, both involving the displacement of magnetic ion $Fe^{2+}$, is attributed to strong spin-phonon coupling [25-28, 31-32, 38-39]. The coupling between the phonons and the spin degrees of freedom can arise either due to modulation of the exchange integral by the phonon amplitude [40-41] and/or by involving change in the Fermi surface by spin waves provided the phonon couples to that part of the Fermi surface [42]. Microscopically, renormalization of the phonon frequency due to spin ordering below the magnetic transition temperature can be correlated [43] with sublattice-magnetization, M(T) as $\Delta\omega \approx [\frac{1}{8\mu_B m\omega}\frac{\partial^2 J}{\partial u^2}]M^2(T)$. Here $\frac{\partial^2 J}{\partial u^2}$ is the second derivative of the spin exchange integral $J$ with respect to the phonon amplitude $u$. To our knowledge there is no report of temperature-dependent sub-lattice magnetization measurement on this system and therefore we could not compare the renormalization of the phonon frequencies with the sub-lattice magnetization.

Since phonon linewidth decreases with decreasing temperature due to reduced anharmonicity, the increase in the phonon linewidths below $T_{sm}$ can not be explained without taking into account the intricate coupling between the phonons and spin degrees of freedom which become prominent below the magnetic transition temperature $T_{sm}$. The sharp increase in the phonon frequencies below $T_{sm}$ clearly suggest the strong coupling between these two degrees of freedom; the spin-phonon coupling will also affect the phonon decay rate below the magnetic transition temperature where the optical phonon may decay into another phonon and magnon/or two magnons, similar to the case of manganites [44].

## 3.2. Orbital-Ordering and Electronic Raman Scattering

As noted earlier, in FeBS iron $d$-orbitals predominantly contribute to the electronic states near the Fermi surface and hence they are expected to play a role in the superconducting pairing mechanism in these systems. In particular, whether the $d_{xz/yz}$ orbitals degeneracy is lifted or not is hotly debated experimentally as well as theoretically [45-53]. We note that a recent report [45] on detwinned Ba(Fe$_{1-x}$Co$_x$)$_2$As$_2$ shows the splitting of Fe $d_{xz/yz}$ orbital much above the $T_{sm}$. Other reports on BaFe$_2$(As$_{1-x}$P$_x$)$_2$ [46, 51-52] and NaFeSe [53] show the existence of electronic nematicity well above the structural and magnetic transition temperature using magnetic torque measurements, $x$-ray absorption spectroscopy and angle resolved photoemission spectroscopy.

Coming to our present study, we observe four weak Raman modes S4 to S7 in addition to the expected first-order Raman modes. The frequencies of the mode S4 is ~ 460 cm$^{-1}$ (45 meV), S5 ~ 620 cm$^{-1}$ (77 meV), S6 ~ 800 cm$^{-1}$ (100 meV) and S7 ~ 1000 cm$^{-1}$ (125 meV). The linewidths of these modes are much higher (~ 5 times) than those of the phonon modes. In conventional Raman scattering intensity of a second-order phonon mode is order of magnitude smaller than its first-order counterpart. In inset of figure 1 we have plotted the intensity ratio of modes S4 to S3, showing that the intensity of mode S4 is much higher than the mode S3, which increases significantly with temperature, ruling out the second-order phononic nature of the mode S4. In earlier Raman studies [22, 54] on FeBS, similar weak and broad Raman modes were observed which were attributed to the electronic Raman scattering involving $d$-orbitals of Fe. We follow the same assignment of the four modes S4 to S7 in terms of the electronic Raman scattering involving Fe $d$-orbitals. Figure 4(a) shows a schematic diagram of Fe$^{2+}$ $d$-levels taken from references [48-50]. The calculated level splitting are close to the frequency position of the high frequency modes. Accordingly, we assign modes S4, S5, S6 and S7 as the transition from the orbital state $z^2$ to orbital states $(x^2-y^2)$, $xz$, $yz$ and $xy$, respectively. Our observation of modes S5 and S6 suggests that the $xz$, $yz$ orbitals are non-degenerate with a split of ~ 25mev, which is close to the theoretical [49-50] and experimental value [45, 52-53] determined using angle resolved photoemission spectroscopy. In Fig. 4(c) and (b) we have plotted the frequencies of modes S5 and S6 and their energy difference ($\delta E$), respectively. It is clearly seen that their splitting ($\delta E$) increases with decreasing temperature, similar to that reported for "122" and "111" systems using angle resolved photoemission spectroscopy [45, 52-53].

Now, we will discuss the origin of high frequency broad Raman band with center near ~ 3200 cm$^{-1}$ as shown in Fig. 5(c). Inset of Fig. 5(c) shows the broad band at room temperature at two different wavelengths i.e. 488 nm and 633 nm showing that the band is not related to photoluminescence, but is a Raman excitation. We have fitted the broad Raman band with a Lorentzian function to extract the peak position and integrated intensity. Figure 5(a) shows the peak position of the broad band as a function of temperature which clearly displays shift of ~ 100 cm$^{-1}$ in the temperature range of 4K to 300K. Figure 5(b) shows the normalized integrated intensity of the broad band showing that $I_{300K}/I_{4K}$ ~ 0.8. We note that similar broad Raman bands in earlier Raman studies on $Fe_{1+y}Se_xTe_{1-x}$ and $BaFe_{2-x}Co_xAs_2$ systems have been attributed to the two-magnon excitations [29-30]. The broad band in ref. [29-30] was observed even at temperature much higher than the magnetic ordering temperature (as high as $5T_{sm}$). Further, in FeBS, even though long range magnetic ordering is not expected in the superconducting phase [8], high frequency Raman band was observed, similar to doped Ba-122 system [29-30]. We assign the broad band near 3200 cm$^{-1}$ as two-magnon excitation. An estimate of the energy scale of the local two magnon excitation in the Heisenberg anti-ferromagnet is $4S(J_{1a} - J_{1b} + 2J_2) - J_{1a}$. Based on experiments, Zhao et al., [55] have deduced the exchange parameters for Ca122 system as $SJ_{1a}$ = 49.9meV, $SJ_{1b}$ = -5.7meV and $SJ_2$ = 18.9meV. Here value of S=1 has been used, as in the case of BaFe$_2$As$_2$ [33]. This value of net spin is less than S=2 because of coupling between various degrees of freedom such as spin-orbit coupling, strong hybridization between Fe-d and As-4p orbitals and itinerant nature of electrons. Using these, the two-magnon energy is ~324 meV (~2600 cm$^{-1}$). Difference between this estimate based on local two-magnon and the observed value can be understood by appreciating that the two-magnon excitation in itinerant Fe 3d anti-ferromagnet is a coupled mode associated with spin and orbital degrees of freedom. This is further strengthened by the fact that the electronic nematic order breaks the $C_4$ rotational symmetry [56] which couples spin and orbital degrees of freedom, as observed in recent experiment [9] as well as theoretical studies [10-15]. A significant intensity of the mode above T$_{sm}$ arises due to short range spin correlations and "coupled" nature of the two-magnon excitation in intinerant antiferromagnet.

## 4. CONCLUSIONS

In conclusion, our temperature dependent Raman study on iron-pnictide Ca(Fe$_{0.97}$Co$_{0.03}$)$_2$As$_2$ provide clear evidence of the strong spin-phonon coupling as reflected in the anomalous temperature of the first-order phonon modes. The four weak modes observed in the range 400-1200 cm$^{-1}$ are attributed to the electronic Raman scattering involving $d$-orbitals of Fe$^{2+}$ and suggest the non-degenerate nature of the $d_{xz/yz}$ orbitals. In addition, the high frequency Raman band is ascribed to two-magnon excitation which has coupled spin and orbital degrees of freedom. Our results suggest that the intricate interplay between spin, charge, lattice and orbital degrees of freedom may be crucial for the pairing mechanism in iron-pnictides. We hope that our experiments will motivate further theoretical work in this direction.

**Acknowledgments**

PK acknowledges CSIR, India, for research fellowship. AKS acknowledge DST, India for financial support. The authors at Dresden thank M. Deutschmann, S. Pichl and J. Werner for technical support and their work was supported by the DFG program FOR 538 and BE1749/13 project.

**FIGURE CAPTION:**

**FIG.1.** (Color online) Raman spectra of $Ca(Fe_{0.97}Co_{0.03})_2As_2$ at 5K. Solid (thin) lines are fit of individual modes and solid (thick) lines show the total fit to the experimental data. Inset shows the integrated intensity ratio of mode S4 with respect to mode S3.

**FIG.2.** (Color online) Temperature evolution of the first-order phonon modes S1 ($A_{1g}$-As), S2 ($B_{1g}$-Fe) and S3 ($E_g$-Fe and As).

**FIG.3.** (Color online) Temperature dependence of the modes S1 ($A_{1g}$-As), S2 ($B_{1g}$-Fe) and S3 ($E_g$-Fe and As). Solid lines are guide to the eyes. Dotted line for mode S1 is fitted curve as described in the text.

**FIG.4.** (color online) (a) Schematic diagram for the crystal-field split energy level of Fe *d*-orbitals showing the electronic transitions. (b), (c) Temperature dependence of the energy difference ($\delta E$) between mode S6 and S5 and their frequency, respectively. Solid lines are guide to the eye.

**FIG.5.** (Color online) (a, b) Temperature dependence of the peak frequency and integrated intensity, respectively. Solid lines are straight line fits. (c) Temperature-dependent Raman spectra in a wide spectral range 1200-5200 $cm^{-1}$. Inset shows the broad Raman band at room temperature at two different wavelength i.e. 488 nm and 633 nm. Solid lines are Lorentzian fits.

FIGURE 1:

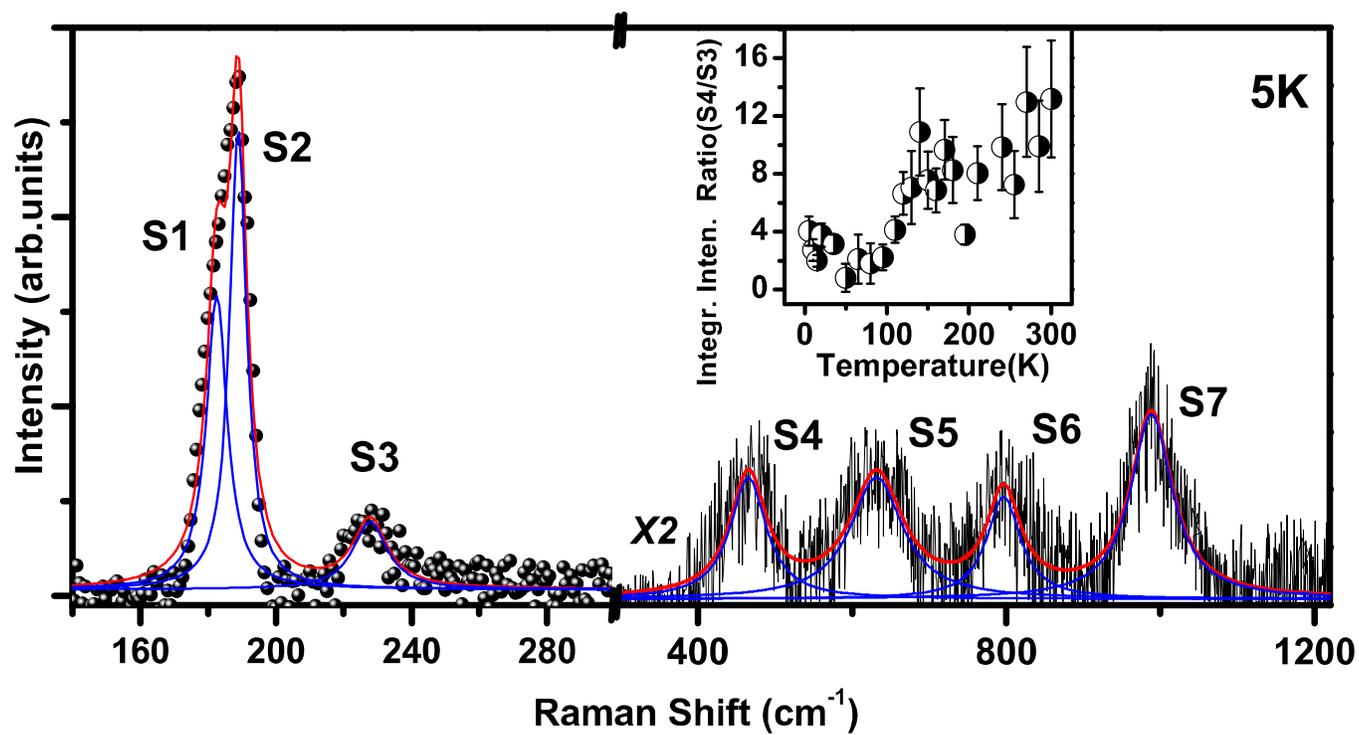

FIGURE 2:

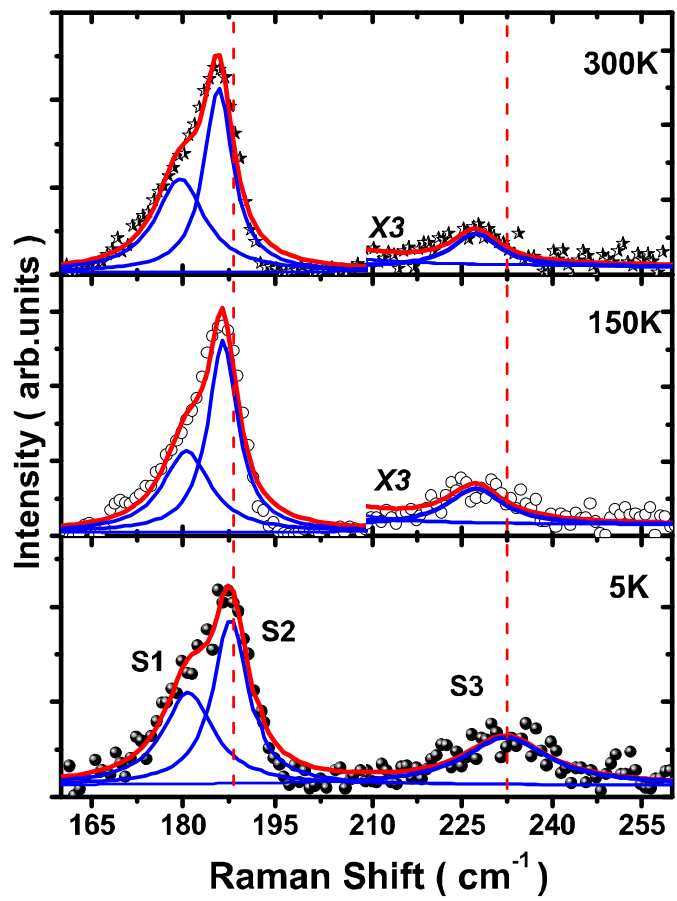

FIGURE 3:

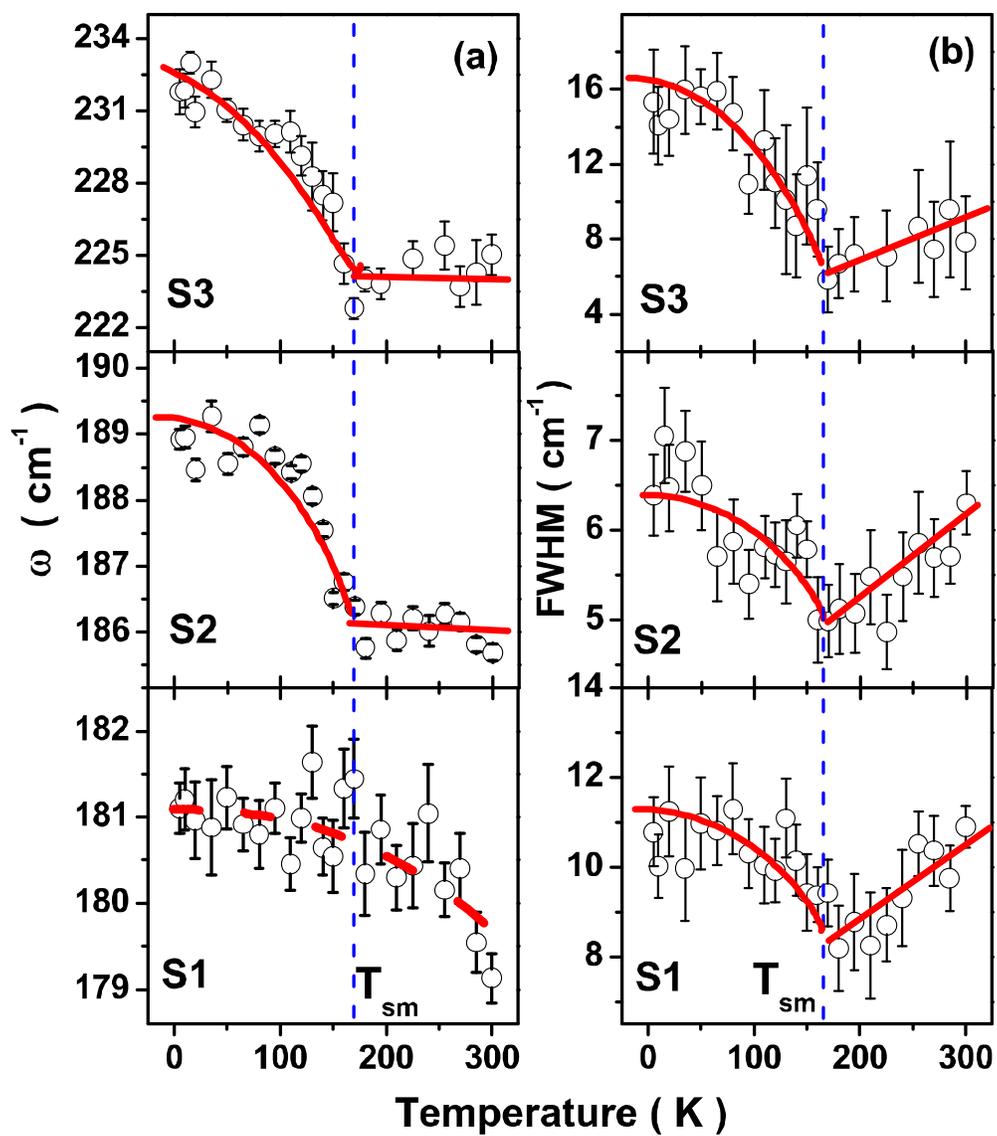

FIGURE 4:

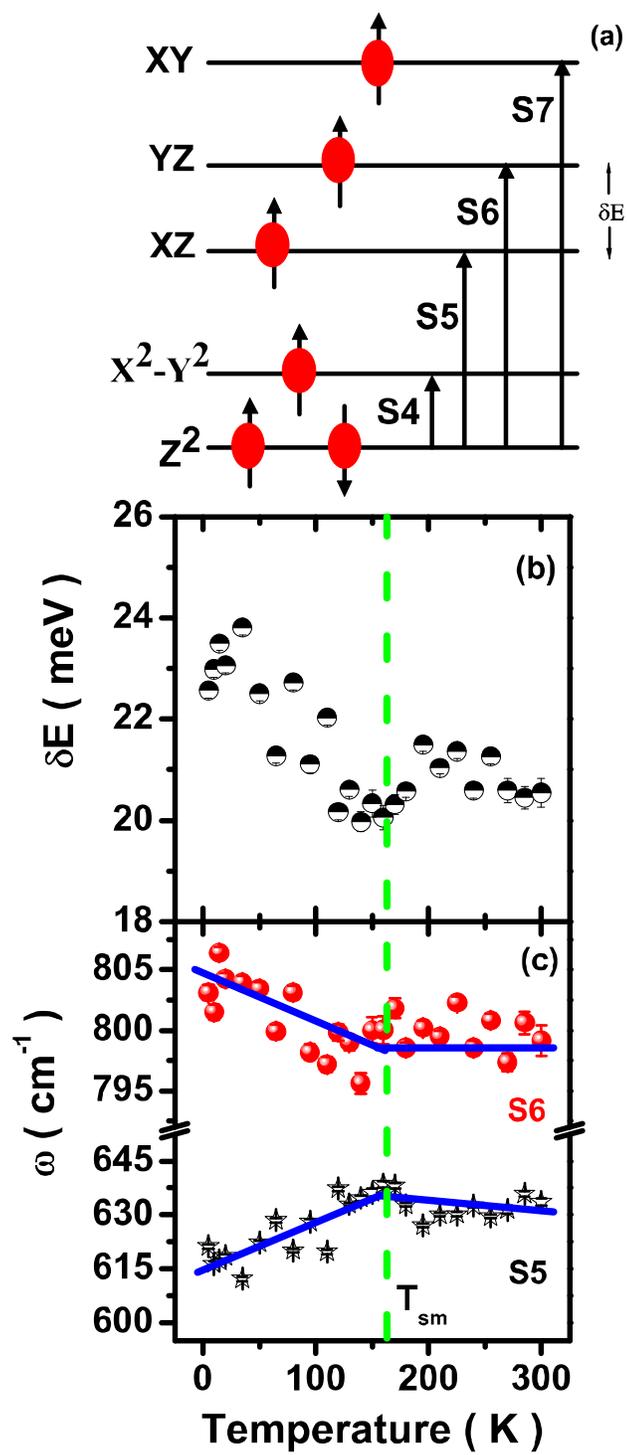

FIGURE 5:

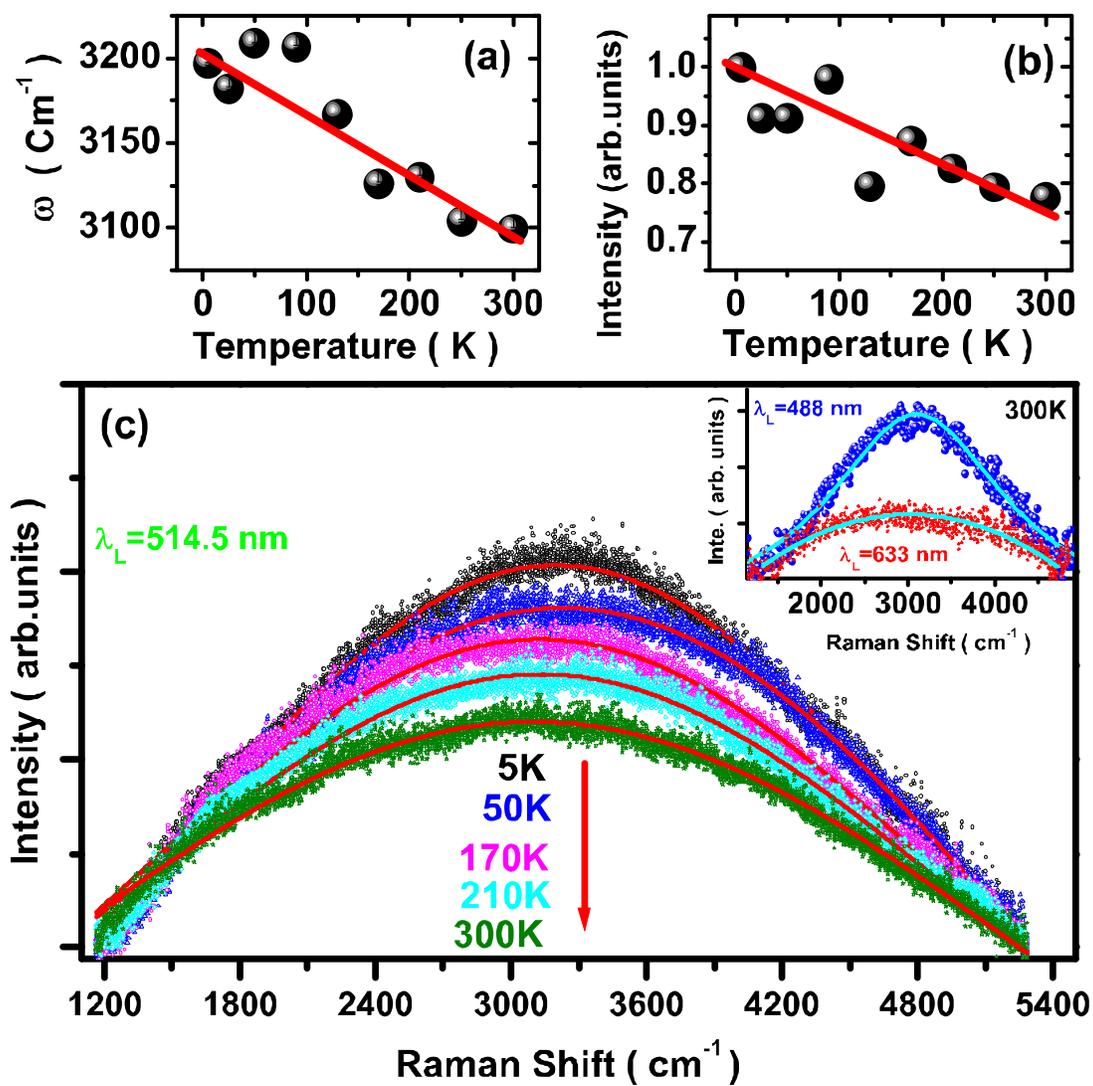